\documentclass[superscriptaddress,aps,twocolumn]{revtex4-2}
\usepackage{comment}
\usepackage[T1]{fontenc}
\usepackage{newtxtext}
\usepackage{newtxmath}

\usepackage{amsmath}
\usepackage{mathtools}
\usepackage{amsfonts}

\usepackage{qcircuit}
\usepackage{braket}

\usepackage{color}
\usepackage{xcolor}
\usepackage[colorlinks,citecolor=blue,linkcolor=blue,urlcolor=blue]{hyperref}

\usepackage{graphicx}
\usepackage[all]{hypcap} 

\usepackage[percent]{overpic}
\usepackage{empheq}
\usepackage[most]{tcolorbox}

\newcommand{\av}[1]{\left\langle #1 \right\rangle }

\newcommand{\Tr}[1]{\text{Tr}\, #1}

\begin{document}

\title{
Impact of Information on Quantum Heat Engines}
\author{Lindsay Bassman Oftelie}
\affiliation{Istituto Nanoscienze-CNR, NEST Scuola Normale Superiore, 56127 Pisa, Italy, EU}
\author{Michele Campisi}
\affiliation{Istituto Nanoscienze-CNR, NEST Scuola Normale Superiore, 56127 Pisa, Italy, EU}

\begin{abstract}
The emerging field of quantum thermodynamics is beginning to reveal the intriguing role that information can play in quantum thermal engines.  Information enters as a resource when considering feedback-controlled thermal machines.  While both a general theory of quantum feedback control as well as specific examples of quantum feedback-controlled engines have been presented, still lacking is a general framework for such machines.  Here, we present a framework for a generic, two-stroke quantum heat engine interacting with $N$ thermal baths and Maxwell's demon.  The demon performs projective measurements on the engine working substance, the outcome of which is recorded in a classical memory, embedded in its own thermal bath.  To perform feedback control, the demon enacts unitary operations on the working substance, conditioned on the recorded outcome.  By considering the compound machine-memory as a hybrid (classical-quantum) standard thermal machine interacting with $N+1$ thermal baths, our framework puts the working substance and memory on equal footing, thereby enabling a comprehensible resolution to Maxwell's paradox and elucidating the intricate manner in which information impacts the performance of quantum engines.  We illustrate the application of our framework with a two-qubit engine. A remarkable observation is that more information does not necessarily result in better thermodynamic performance: sometimes knowing less is better.

\end{abstract}

\maketitle

\section{Introduction}
Quantum thermodynamics, which aims to elucidate the fundamental principles of thermodynamics in the quantum regime, has progressed rapidly over the last couple of decades \cite{Gemmer09Book,Binder18Book,deffner2019quantum,Strasberg22Book,campbell2025roadmap}.  Spurred in part by the miniaturization of technological devices, the development of quantum thermodynamics is crucial for improving design and performance of microscopic heat engines and refrigerators \cite{quan2007quantum, denzler2020efficiency, solfanelli2023quantum, levy2012quantum, brunner2014entanglement, kosloff2014quantum, campisi2015nonequilibrium, tajima2021superconducting, kamimura2022quantum, cangemi2024quantum}. Of particular interest is the role that information plays in such systems, an avenue that is currently being explored within the framework of feedback-controlled thermal machines. Such machines harness information gained via measurement of the working substance (WS) to steer the evolution of the system, potentially providing performance enhancement or even enabling processes that would otherwise be forbidden \cite{park2013heat, brandner2015coherence, elouard2018efficient}. 

Work along this theme began with the famous thought experiment of Maxwell's demon \cite{maxwell1871theory}, an intelligent being that seemed able to violate the second law of thermodynamics by utilizing information to reduce entropy in a closed system.  The concept of information as a quantifiable resource in thermodynamic processes was further elucidated with the introduction of Szilard's engine, an engine fueled by information \cite{szilard1929entropieverminderung}. Eventually, the demon was exorcised, restoring the second law of thermodynamics, by recognizing that writing and erasure of information in the demon's memory must be taken into account when calculating the total entropy change in the system \cite{ landauer1961irreversibility, bennett1982thermodynamics}.  There have been continued efforts to better understand the nature and role of information in thermodynamics processes in the quantum regime \cite{Loyd97PRA56, kieu2004second, quan2006maxwell, Sagawa08PRL100, sagawa2009minimal, sagawa2012thermodynamics, mandal2013maxwell, parrondo2015thermodynamics, goold2016role}, while a number of experimental works have demonstrated the use of information in quantum thermal engines \cite{toyabe2010experimental, koski2014experimental, camati2016experimental, cottet2017observing, masuyama2018information, kumar2018sorting, aggarwal2025rapid}.  However, a complete understanding of the impact of information in thermal engines is still under active investigation (see Ref. \cite{Oliveira25PRXQ6} for a review, in particular Section VIII C). 

While there are specific examples of feedback-controlled quantum heat engines in the literature, a general framework has yet to be presented. Meanwhile, the highly abstracted theory of quantum feedback control by Sagawa and Ueda \cite{Sagawa08PRL100} is so general that its application to quantum heat engines is not immediately apparent. In this paper, we aim to bridge this gap between explicit examples and general theory by grounding feedback-controlled quantum heat engines in a more physically tangible framework. 


\begin{figure}\label{fig:schematic}
\centering
\begin{overpic}[width=0.74\columnwidth]{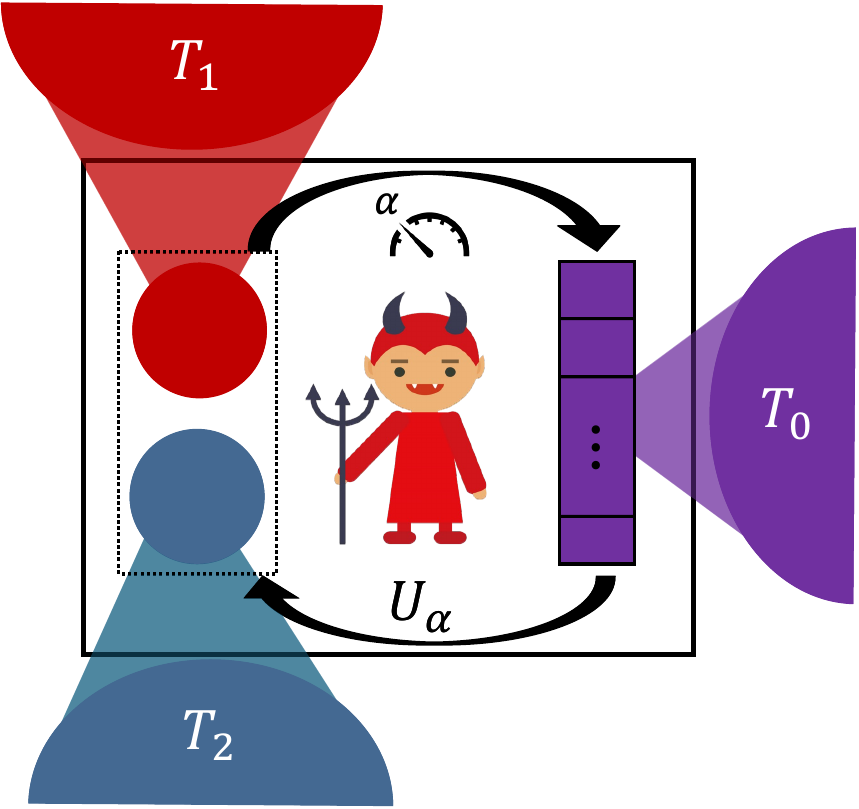}
\end{overpic}
\caption{Schematic of the physical model for a feedback-controlled quantum heat engine. The multi-partite working substance (WS) of the heat engine, outlined in a dotted black rectangle, is comprised of quantum systems, each in contact with their own thermal bath, possibly at different temperatures, $T_1, ..., T_N$.  The extended working substance (xWS), outlined by the solid black rectangle, extends the heat engine to include the demon's memory, which is in thermal contact with yet another thermal bath at temperature $T_0$.  The demon measures the WS, writing the result $\alpha$ into memory.  Based on the value of $\alpha$, the demon enacts a corresponding unitary $U_{\alpha}$ on the WS.}
\end{figure}

Specifically, we consider a multi-bath system, depicted schematically in Figure  \ref{fig:schematic}, comprising a two-stroke heat engine, which undergoes measurement and feedback control, combined with a classical memory that stores the information collected from the measurement.  The heat engine itself is comprised of a bipartite (or in general, a multi-partite) WS, in which each component can be independently coupled/decoupled to its own thermal bath at a distinct temperature.  Likewise, the classical memory is also coupled to its own thermal bath.  We highlight that an advantage of our framework is that it treats the WS and the demon's memory on equal footing, removing some of the often perceived mysteriousness of the demon's actions.  This enables us to derive what we believe to be a particularly comprehensible and accessible exorcism of Maxwell's demon, thereby rigorously restoring the second law of thermodynamics within our model.

We next leverage our model to explore the efficiency of feedback-controlled quantum heat engines. The large number of free parameters in the system presents a Pandora's box of scenarios to study. However, we narrow our focus to a few relevant regimes of initial conditions and types of projective measurements to uncover several intriguing observations. In particular, we prove that for fine-grain measurements (which we define more formally later), there exists an optimal measurement basis that maximizes the engine's efficiency. In addition, and perhaps counter-intuitively, we find that in some scenarios, coarse-grain measurements can lead to greater efficiency than fine-grain measurements; in other words, sometimes knowing less is better. Our findings provide valuable insights into the interplay of information and energy conversion at the quantum scale, offering guidance for the design of practical quantum thermal devices.

\section{Theory}
\subsection{The working substance}
We consider a multi-partite quantum system as the WS of the thermal machine. Let 
\begin{align}
H= \sum_{i=1}^N H_i
\label{eq:H}
\end{align}
denote the WS Hamiltonian, with $H_i$ the Hamiltonian of each partition thereof.
For the sake of simplicity we restrict the discussion to the so-called two-stroke cycle \cite{quan2007quantum,campisi2015nonequilibrium}. In the two-stroke cycle, the WS begins in a multi-partite Gibbs state $\rho$, featuring each partition at thermal equilibrium with possibly distinct temperatures, written as
\begin{align}
\rho = \bigotimes_{i=1}^N \frac{e^{-\beta_i H_i}}{Z_i},\quad Z_i= \Tr e^{-\beta_i H_i}.
\label{eq:rho}
\end{align}

\subsection{The first stroke}
The first stroke begins with the WS undergoing a projective measurement of some observable 
\begin{align}
A = \sum_{\alpha=1}^K a_\alpha \Pi_\alpha  
\label{eq:A}
\end{align}
where $a_\alpha $ denote its eigenvalues, and $\Pi_\alpha$ denote the corresponding eigenprojectors. We shall denote with $g_\alpha$ the rank of projector $\Pi_\alpha$, that is the degeneracy of the eigenvalue $a_\alpha$. Here, $K$ is the number of distinct eigenvalues of $A$, which is bounded by the WS Hilbert space dimension $d= \sum_{\alpha=1}^Kg_\alpha$.
As a consequence of the measurement, one outcome, say outcome $\alpha$, is produced and recorded in a classical memory. According to the postulates of quantum mechanics, the state $\rho$ collapses onto 
\begin{align}
\rho_\alpha =  \frac{\Pi_\alpha \rho \Pi_\alpha}{q_\alpha}
\end{align}
where
\begin{align}
q_\alpha =\Tr\, \Pi_\alpha \rho \Pi_\alpha
\label{eq:qalpha}
\end{align}
is the probability that the $\alpha$-th outcome is produced. Note that the effect of the measurement on the WS is fully specified  by the set of measurement projectors
\begin{align}
    \boldsymbol{\Pi} = \{\Pi_1, \dots \Pi_K\}.
\end{align}
The eigenvalues $a_\alpha$ do not play any role, hence in the following they will not be mentioned any more.

The demon has a collection of $K$ unitaries, $U_\alpha$, $\alpha=1 \dots K$, and feedback control consists of the demon enacting unitary $U_\alpha$ on the WS, conditioned on the outcome $\alpha$ being measured. The post-feedback-control WS state then reads:
\begin{align}
\rho'_\alpha = U_\alpha \rho_\alpha U^\dagger_\alpha
\label{eq:rho'alpha}
\end{align}
With this operation, the first stroke is completed.

Two points are worth stressing. First, during the first stroke the WS does not interact with any thermal bath.  Second, in order to enact the unitary $U_\alpha$ the demon has to implement a cyclic time dependent perturbation $V_\alpha(t)$, such that the resulting WS Hamilton function $H_\alpha(t)=H+V_\alpha(t)$ induces the desired unitary $U_\alpha$. That is, $V_\alpha$ must be such that
$
U_\alpha= T\exp[-i/\hbar \int_0^\tau ds H_\alpha(s)]
$. 
Typically $V_\alpha(t)$ will contain interaction terms among the various parts of the WS. In the following we shall only focus on the unitaries $U_\alpha$ without entering the separate problem of designing the according $V_\alpha$.

\subsection{The second stroke}
The second stroke consists of placing each partition of the WS in weak thermal contact with its associated thermal bath so that it reaches a state of thermal equilibrium and leads the WS back to the initial state Eq. (\ref{eq:rho}), thereby closing the cycle.

\subsection{Apparent violation of the second law}
In order to access the compliance of our engine  model with the second law of thermodynamics, we must first compute the change in the expectation value of the energy of each partition $i$ across the first stroke, reading
\begin{align}
\av{\Delta E_i} = \Tr H_i (\rho'-\rho)
\label{eq:delta_E}
\end{align}
where
\begin{align}
\rho' = \sum_{\alpha=1}^K q_\alpha \rho'_\alpha = \sum_{\alpha=1}^K  U_\alpha \Pi_\alpha \rho \Pi_\alpha  U^\dagger_\alpha = \mathcal M [\rho]
\label{eq:rho'}
\end{align}
is the non-post-selected WS state at the end of the first stroke.
Note that we denote with $\mathcal M$ the linear transformation that maps $\rho$ onto $\rho'$.

As shown in Ref. \cite{Campisi17NJP19}, due to the special initial condition in Eq. (\ref{eq:rho}), the following inequality holds:
\begin{align}
\sum_{i=1}^N \beta_i \av{\Delta E_i}  \geq \Delta \mathcal H = \mathcal H[\rho'] -\mathcal H[\rho]
\label{eq:clausius0}
\end{align}
where $\mathcal H$ denotes von Neumann information:
\begin{align}
\mathcal H[\sigma] = -\Tr \sigma \ln \sigma.
\end{align}

Depending on how the unitaries $U_\alpha$ are selected, the von Neumann information may increase, decrease, or remain constant. For example, if one and the same unitary is enacted regardless of the outcomes (i.e., no feedback control), that is $U_\alpha=U ~~\forall ~ \alpha$, the resulting map $\mathcal M$ would be unital, thus resulting in an overall increase in von Neumann information \footnote{To see this, use the resolution of the identity, $ \sum_{\alpha=1}^K \Pi_\alpha= \mathbb{I}$, and the unitarity of $U$, implying $U U ^\dagger= \mathbb{I}$}. Therefore, active feedback control is necessary for compressing the von Neumann information \cite{Campisi17NJP19}.

Assuming the system-bath interaction energy is so weak that it can be neglected, the negative energy change $\av{\Delta E_i}$ in each partition of the WS across the first stroke, equals the average heat $Q_i$ ceded by thermal bath $i$ to the WS during the second stroke
\begin{align}
Q_i =  -\av{\Delta E_i}.
\label{eq:Q_i}
\end{align}
Thus, during the second stroke, the energies $\av{\Delta E_i}$'s that the parts of the WS recieve via interaction with the work source and the other parts during the first stroke, are distributed to the thermal baths.

Combining Eq. (\ref{eq:Q_i}) with Eq. (\ref{eq:clausius0}), we have 
\begin{align}
\sum_{i=1}^N \beta_i Q_i  \leq - \Delta \mathcal H. 
\end{align}
When $\mathcal M$ induces a compression of von Neumann information $\Delta \mathcal H <0$, it is possible that the Clausius sum $\sum_{i=1}^N \beta_i Q_i $ takes a positive value.  In this case, it would \emph{appear} that the second law of thermodynamics has been violated in a way that is completely analogous to Maxwell's demon paradox, except here the WS is a quantum system.

\subsection{Reconciling the second law, exorcism of Maxwell's demon}
As is well-known, resolving the paradox of the apparent violation of the second law of thermodynamics requires taking into account the thermodynamic cost of the demon's information gathering.  As widely discussed in the literature, see e.g., \cite{Vaikuntanathan11PRE83}, due to the Landauer principle \cite{Landauer61IBMRD5}, the average energy, $\av{W}_\text{m}$,  spent by the demon to gather information, i.e., to imprint the outcome labels $\alpha$ in the classical memory, obeys the bound:
\begin{align}
    \av{W}_\text{m} \geq \beta_0^{-1} h[\{q_\alpha\}]
    \label{eq:av-m>h}
\end{align}
where 
\begin{align}
 h[\{q_\alpha\}] \doteq - \sum_{\alpha=1}^K q_\alpha \ln q_\alpha
\end{align}
is the Shannon information of the distribution $\{q_\alpha\}$, where $q_\alpha$ is defined in Eq. (\ref{eq:qalpha}), and $\beta_0$ is the inverse thermal energy of the bath in which the classical memory is embedded.  

We stress that such an encoding process is nothing but the quantum measurement itself. In fact, a quantum measurement process is the process of imprinting information about the state of a quantum system into a large classical system (i.e., the classical memory), such that it can be read by a classical being (i.e., the demon).  Accordingly, $\av{W}_\text{m}$ can be understood as the energetic cost  of the quantum measurement.

It is important to clarify that with ``imprinting information'' in the memory, we mean setting the memory to the specific state $\alpha$ (corresponding to the state of the measured quantum system). Physically, such a process of ``RESET to $\alpha$'' is equivalent to what Landauer calls an ``erasure'', so its minimal energetic cost is, in fact, given by Eq. (\ref{eq:av-m>h}), which is universally known as the cost of ``erasure'', rather than ``imprinting'' \cite{Landauer61IBMRD5}. Unfortunately, this nomenclature can raise some confusion, and we hope to have clarified the fact that ``erasure'' and ``imprinting'' are just two sides the same coin. It is also worth remarking that, in our model, the memory need not be reset to any special state after imprinting the label $\alpha$, because the next measurement will encode the new state $\alpha'$ regardless of the memory previous state $\alpha$, which is the defining feature of the ``RESET to'' process \cite{Landauer61IBMRD5,Buffoni22JSP186}.

%
%

A crucial result is the following inequality
\begin{align}
h[\{q_\alpha\}] \geq -\Delta \mathcal H ,
\label{h>DH}
\end{align}
which implies that the negative von Neumann information change in the WS is smaller than the Shannon information of the measurement outcomes.

A proof of Eq. (\ref{h>DH}) can be found in Appendix \ref{sec:proof}, while here we discuss its most crucial consequence.
Using Eq. (\ref{eq:av-m>h}), we get
\begin{align}
\av{W}_\text{m} \geq  -\beta_0^{-1} \Delta \mathcal H
\end{align}
which, when combined with Eq. (\ref{eq:clausius0}), leads to the salient result
\begin{align}
 \beta_0 \av{W}_\text{m} + \sum_{i=1}^N \beta_i \av{\Delta E_i} \geq 0
\end{align}
or, equivalently 
\begin{align}
\label{eq:xClausius}
\sum_{i=0}^{N} \beta_i Q_i \leq 0
\end{align}
where we introduce the notation
\begin{align}
Q_0=-\av{W}_\text{m}  
\label{Wm=-Q0}
\end{align}
to denote the average heat ceded by the thermal bath coupled to the memory.
Note that the sum in Eq. (\ref{eq:xClausius}) runs from $i=0$, and hence it includes the heat $Q_0$.

Equation (\ref{eq:xClausius}) expresses the validity of the Clausius inequality (i.e., the second law of thermodynamics) for the hybrid (classical-quantum) extended WS (xWS) comprising the $N$-partite quantum WS and the classical memory. 
This crucial inequality allows us to treat the extended hybrid engine as a standard feedback-less heat engine comprising $N+1$ thermal baths.  Heat is exchanged between the WS and the thermal baths, while work is exchanged between an external agent that writes measurement outcomes into a classical memory (in contact with one of the thermal baths) and enacts the $K$ unitaries $U_\alpha$ on the WS (in contact with the remaining $N$ thermal baths).

Since, at the end of the two-stroke thermodynamic cycle, the WS  returns to its initial state, assuming all memory states have same energy, the total work output equals the total energy ceded by the baths.  Thus, Eq. (\ref{eq:xClausius}) is complemented by the following equation 
\begin{align}
W_\text{out} = \sum_{i=0}^{N} Q_i,
\label{eq:x1st}
\end{align}
which expresses the first law of thermodynamics for the xWS. Here, $W_\text{out}$ denotes the total work output of the xWS, which can be rewritten as

\begin{align}
W_\text{out} = \av{W}^\text{out}_\text{fc} - \av{W}_\text{m}.
\label{eq:W-out}
\end{align}

where, $\av{W}^\text{out}_\text{fc}$, is the work extracted from the WS across the first stroke by enacting the feedback control unitaries $U_\alpha$, namely, 
\begin{align}
\av{W}^\text{out}_\text{fc}= -\sum_{i=1}^N \av{\Delta E_i}  = \Tr H (\rho-\rho')\, .
\label{eq:w-fc}
\end{align}

Note that the $W_\text{out}$ is the difference of two competing terms. The quantity $\av{W}^\text{out}_\text{fc}$ is the energy output from the WS during the first stroke. The quantity $\av{W}_\text{m}$ is the energy input to perform measurements. 
We stress that the cost of measurement $\av{W}_\text{m}$ depends on measurement projectors $\boldsymbol{\Pi}=(\Pi_1, \dots \Pi_K)$ but not on the according unitaries $\boldsymbol{U}=(U_1, \dots U_K)$.
In contrast, given the projectors $\boldsymbol{\Pi}$, the energy the demon gets out of the WS, $\av{W}^\text{out}_\text{fc}$, depends on its choice the unitaries $\boldsymbol{U}$. 

As will be discussed in more detail below, in Sec. \ref{sec:max-work-L-eff} and  illustrated with the examples in Sec. \ref{sec:examples}, this gives rise to a highly
complex trade-off between the utility of the information and
the cost to retrieve it.


Note that in the case of a single heat bath (that thermalizes both the WS and memory), Eq. (\ref{eq:xClausius}) boils down to:
\begin{align}
W_\text{out} \leq 0
\end{align}
expressing the Kelvin postulate of the second law: no energy can be extracted from a single thermal bath by the application of a cyclical driving. This generalizes the classical result of Ref. \cite{Vaikuntanathan11PRE83} to the quantum case.

\section{Efficiency of heat engines}

Depending on the relative signs of $W_\text{out}$ and $Q_i$, distinct operation modes of the engine can be realized \cite{Buffoni19PRL122,Solfanelli20PRB101}. The most relevant are the ``heat engine'' mode, characterized by positive work output, and the ``refrigeration mode'', characterized by positive heat extraction from the lowest temperature bath.
In the following, we shall focus on the heat engine operation, i.e., we shall assume $W_\text{out}>0$, which in turn implies that we are considering the case in which there are at least two distinct temperatures $T_i$.

In this regime the thermodynamic efficiency is given by
\begin{align}
\eta= \frac{W_\text{out}}{Q_\text{in}} 
\end{align}
where $Q_\text{in}$ is the total heat intake 
per cycle
\begin{align}
Q_\text{in} = \sum_{i=0}^N [Q_i]_+
\label{eq:Qin}
\end{align}
where 
$[x]_+=x \theta(x)$
with $\theta$ denoting Heaviside step function. In other words, $Q_{\text{in}}$ is given by the sum of all positive values $Q_i$.

According to Ref. \cite{Campisi16JPA49a}, Eqs. (\ref{eq:xClausius}, \ref{eq:x1st}) imply that 
$\eta$ is bounded by the Carnot efficiency $\eta^C$
\footnote{Similarly, in the case when the hybrid machine operates as a refrigerator, its cooling efficiency is bounded by the cooling Carnot efficiency \cite{Campisi16JPA49a}}:
\begin{align}
\eta \leq 1- \frac{T_\text{min}}{T_\text{max}}\doteq \eta^C
\label{eq:eta}
\end{align}
where $\eta^C$ is expressed in terms of the largest temperature $T_\text{max}$ of all the baths that give away heat, and the smallest temperature $T_\text{min}$ of all the baths that receive heat
\footnote{For simplicity here we adopt the convention that temperature is measured in units of energy, which is formally equivalent to setting Boltzmann constant $k_B$ to unity}, 
\begin{align}
T_\text{min} = \min_{\{i|Q_i<0\}} \beta_i^{-1}, \quad T_\text{max} = \max_{\{i|Q_i>0\}}\beta_i^{-1} \,. 
\end{align}

In the context of our xWS, $Q_{\text{in}}$ is the sum of the individual heat intakes of each of the $N+1$ thermal baths, Eq. (\ref{eq:Qin}).  Recall that index $i=0$ denotes the memory bath, while $i=1 \dots N$ denote the thermal baths associated with the WS. In general, according to Eqs. (\ref{eq:av-m>h},\ref{Wm=-Q0}) the heat ceded by the memory bath
\begin{align}
    Q_0 \leq -\beta_0^{-1}h[\{q_\alpha\}]\leq 0
\end{align}
because the Shannon information is non-negative. This inequality reflects the fact that in the present set-up, the memory bath always receives a positive heat since the process of memory imprinting is dissipative.
Thus, the total heat intake can be rewritten 
 \begin{align}
Q_\text{in} = \sum_{i=1}^N [Q_i]_+
\label{eq:Q-in}
\end{align}
where the sum now runs from $i=1$.

Using Eq. (\ref{eq:x1st}), we have 
\begin{align}
    \eta = \frac{\sum_{i=1}^NQ_i}{Q_\text{in}} + \frac{Q_0}{Q_\text{in}}.
\end{align}
In general, $Q_0$ will depend on the intricacies of how the demon performs its information gathering.  However, the Landauer bound, Eq. (\ref{eq:av-m>h}) sets a maximum for its value as
\begin{align}
    Q_0^L= -T_0 h[\{q_\alpha\}]
    \label{eq:Q0L}
\end{align}
where $T_0=\beta_0^{-1}$ is the temperature of the memory bath.  Therefore, in the following, we shall focus on what we will call the ``Landauer efficiency''
\begin{align}
\eta^L = \frac{\sum_{i=1}^NQ_i}{Q_\text{in} } -  \frac{h[\{q_\alpha\}]} {Q_\text{in} }T_0
\label{eq:eta-L}
\end{align}
which denotes the efficiency achieved when the Landauer bound is saturated, i.e., when $Q_0$ takes on its maximal value. We emphasize, however, that in any realistic realisation of the engine, the actual efficiency $\eta$ will be smaller than $\eta^L$,which in turn is smaller than Carnot efficiency $\eta^C$:
\begin{align}
    \eta \leq \eta^L \leq \eta^C\, .
\end{align}

Note that the cost of information gathering only enters $\eta^L$ linearly through $T_0$.  The smaller $T_0$ the higher $\eta^L$, reflecting that less dissipation is incurred gathering in information at lower memory bath temperature.

\section{Maximum Landauer efficiency}
It is interesting to analyze the maximum Landauer efficiency that can be achieved in feedback-controlled quantum heat engines. For fixed WS Hamiltonian $H$ and bath temperatures, 
$\boldsymbol{\beta}=(\beta_0, \beta_1, \dots \beta_N)$, the Landauer efficiency $\eta^L$ depends on the measurement projectors $\boldsymbol{\Pi}=(\Pi_1, \dots \Pi_K)$ and the associated feedback unitaries $\boldsymbol{U}=(U_1, \dots U_K)$. In particular, for a given set of measurement projectors $\boldsymbol{\Pi}$, $\eta^L$ changes depending on the choice of unitaries $\boldsymbol{U}$. In the following, we shall use the notation $\bar\eta^L$ to denote the maximum of $\eta^L$ over all possible choices of feedback unitaries $\boldsymbol{U}$. That is (leaving the dependence on $H,\boldsymbol{\beta}$ implicit, in order to keep the notation light):
\begin{align}
    \bar\eta^L(\boldsymbol{\Pi}) \doteq \max_{\boldsymbol{U}} \eta^L(\boldsymbol{\Pi},\boldsymbol{U})
\end{align}
For simplicity, we shall call it the ``max Landauer efficiency''.

In order to analyze the max Landauer efficiency, it is helpful to distinguish between two main paradigms of measurement: fine-grain and coarse-grain measurement.  Fine-grain measurement refers to measurements in which the rank of all the measurement projectors $\Pi_\alpha$ is unity. 
A coarse-grained measurement refers to any measurement in which at least one projector has rank larger than $1$.



\subsection{Fine-grain measurement}
It is convenient to introduce the spectral decomposition of the Hamiltonian
\begin{equation}
    H = \sum_{k=1}^d e_k \ket{e_k}\bra{e_k}
\end{equation}
in terms of its ordered eigenvalues $e_1 \leq e_2 \dots \leq e_d$ and eigenvectors $\ket{e_k}$. 
We shall use the symbol 
\begin{align}
    \boldsymbol{E}=(\ket{e_1}\bra{e_1},\dots ,\ket{e_d}\bra{e_d})
    \label{energy_basis}
\end{align}
to denote the complete set of fine-grain measurement operators of $H$.

A first remarkable result is the following: given a generic set of fine-grain measurement operators, $\boldsymbol{F}=(\ket{f_1}\bra{f_1},\dots ,\ket{f_d}\bra{f_d}))$, where $(\ket{f_1},\dots ,\ket{f_d}$) form a complete orthonormal basis of the WS Hilbert space, we have
\begin{align}
    \bar\eta^L(\mathbf{F}) = 1 - \frac{h[\{q_k\}]} {\bar Q_\text{in} }T_0
    \label{eq:top-eta-L-fine}
\end{align}
where $q_k = \bra{f_k}\rho \ket{f_k}$ is the probability that the $k$-th outcome of the measurement is produced, and 
\begin{align}
    \bar Q_\text{in} = \Tr H (\rho - \ket{e_1} \bra{e_1})
    \label{eq:Q-bar-in}
\end{align}
is the difference between the WS ground state energy and its initial energy.


To prove Eq. (\ref{eq:top-eta-L-fine}), note that $\eta^L$, Eq. (\ref{eq:eta-L}), is composed of two terms. The first term $\sum_{i=1}^N Q_i/ \sum_{i=1}^N [Q_i]_+$ is smaller than or equal to $1$, due to the inequality $x\leq [x]_+$. Its maximum value, i.e., $1$, is obtained when all $Q_{i}$'s (with $i>0$) are positive. The second term $h[\{q_k\}]/Q_\text{in}$ is minimized when $Q_\text{in}$ is maximum (note that the numerator does not depend on the unitaries $U_k$). Such maximum occurs when the unitaries $U_k$ are the unitaries $\bar U_k $ that extract the highest amount of energy (i.e., the ergotropy \cite{Allahverdyan04EPL67}) from the post-selected states $\rho_k$. 
We refer to these unitaries as \textit{ergotropy unitaries}, which in general, send the post-selected states $\rho_k$ to their passive companion states $\breve \rho_k $.

Since the post-measurement states $\rho_k= \ket{f_k}\bra{f_k}$ of fine-grain measurement are pure, 
their passive companion is the ground state $\ket{e_1}\bra{e_1}$, for all $k$'s and the ergotropy unitaries for fine-grain measurement satisfy
\begin{align}
    \bar U_k \ket{f_k} = e^{-i\phi_k }\ket{e_1}, ~~~ k=1 \dots d.
    \label{eq:ergotropyU}
\end{align}
Thus, when such unitaries $\bar U_k$ are chosen for feedback control, the post-feedback-control state, Eq. (\ref{eq:rho'}), is the WS Hamiltonian ground state,  $\rho'=\ket{e_1} \bra{e_1}$. Given that the WS  Hamiltonian is the direct sum of the $N$ local Hamiltonians $H_i$, the ground state is the tensor product of the ground states of each partition. Thus, the unitaries $\bar U_k$  individually maximize each term $Q_{i}$ (with $i>0$).  Since all such maximum $Q_i$'s are trivially non-negative, their sum amounts to $Q_\text{in}$, Eq. (\ref{eq:Q-in}), which thus attains its maximal value $\bar Q_\text{in}$ in Eq. (\ref{eq:Q-bar-in}). 

Note that this value is one and the same regardless of the set of measurement projectors, as long as they are all rank 1. Furthermore, the positivity of all $Q_i$'s (with $i>0$) also ensures maximization of the first term $\sum_{i=1}^N Q_i/ \sum_{i=1}^N[Q_i]_+$, which attains unit value. Accordingly, the difference of the two terms in Eq. (\ref{eq:eta-L}), i.e., the Landauer efficiency $\eta^L$, is maximized by the ergotropy unitaries, Eq. (\ref{eq:ergotropyU}), and amounts to the value in Eq. (\ref{eq:top-eta-L-fine}).



A remarkable aftermath of Eq. (\ref{eq:top-eta-L-fine}) is that, among all fine-grain measurement sets, the largest value of the max Landauer efficiency is achieved when measuring in the WS Hamiltonian eigenbasis $\boldsymbol{E}$. That is,
\begin{align}
    \max_{\boldsymbol{F}} \bar\eta^L(\boldsymbol{F}) = \bar\eta^L(\boldsymbol{E}).
    \label{eq:max-L-eff}
\end{align}

To prove Eq. (\ref{eq:max-L-eff}) note that the measurement operators enter Eq. (\ref{eq:top-eta-L-fine}) only through the Shannon information. Thus, we only need to show that the Shannon information $h[\{\cdot\}]$ is minimal among all fine-grain measurements when measuring over the set $\boldsymbol{E}$. Let
\begin{align}
    \rho = \sum_{n=1}^d p_n \ket{e_n}\bra{e_n}
    \label{eq:rho-spectral}
\end{align}
be the spectral decomposition of the initial state (recall, that by construction, $\rho$, Eq. (\ref{eq:rho}), is diagonal in the WS Hamiltonian eigenbasis).
Using Eq. (\ref{eq:rho-spectral}), we get
\begin{align}
    q_k = \bra{f_k}\rho \ket{f_k} = \sum_{n=1}^d p_n \braket{f_k|e_n}\braket{e_n|f_k}.
\end{align}
Note that $P_{kn}\doteq \braket{f_k|e_n}\braket{e_n|f_k}=|\braket{f_k|e_n}|^2$ are the elements of a bistochastic matrix, therefore \cite{Sagawa20Book}:
\begin{align}
    h[\{q_k\}] = h[\{\sum_n P_{kn} p_n \}] \geq h[\{p_k\}]
    \label{eq:E-is-best}
\end{align}
Thus, the fine-grain measurement over the Hamiltonian eigenbasis $\boldsymbol{E}$ is the one that collects the smallest amount of information, hence dissipates the least amount of heat in the memory bath, and thus achieves the largest max Landauer efficiency.

\subsection{Coarse-grain measurement}
The picture gets considerably more complicated when coarse-grain measurements are considered. As will become clear with the illustrative examples below, in this case the ergotropy unitaries $\bar U_\alpha, \alpha =1  \dots K<d$ do not all send the state of the system to one and the same state $\rho'_\alpha$ as in the case of fine-grain measurement, nor generally extract energy from all baths $i=1 \dots N$. 
Thus, Eq. (\ref{eq:top-eta-L-fine}) does not apply in this case, nor does it appear plausible that a similar simple analytical formula can be found. Depending on the specific set up, it can be difficult to find the set of unitaries that maximize $\eta^L$ due to its non-linearity, and the problem should be generally treated numerically.

It is important to remark that using coarse-grain measurements is a way to collect less information, thereby dissipating less heat into the memory bath and increasing efficiency. Less information, however, can negatively impact heat extraction from the other baths, thereby reducing work output and decreasing efficiency. Thus, there is a trade-off between the cost of gathering information and the gain one can derive from it. As we shall see in the examples below, sometimes knowing less can be advantageous.

\section{Landauer efficiency at maximum work output}
\label{sec:max-work-L-eff}
Depending on the application, it may be more desireable to maximize work output, rather than the efficiency when designing a heat engine.  Therefore, we now turn our attention to the Landauer efficiency at maximum work output, which we denote by 
$
    \eta^{L}_{W}(\boldsymbol{\Pi})
$, and call it the ``max-work Landauer efficiency''. It is similar to the ``efficiency at maximum power'' of  Curzon and Ahlborn \cite{Curzon75AJP43}.  It is defined as the value that the Landauer efficiency $\eta^L(\boldsymbol{\Pi},\boldsymbol{U})$, Eq. (\ref{eq:eta-L}), attains when the unitaries $\boldsymbol{U}$ are chosen to maximize the work output. 

Note that the unitaries only enter $W_\text{out}$, Eq. (\ref{eq:W-out}), through the term $\av{W}^\text{out}_\text{fc}$. Thus, maximizing $W_\text{out}$ is equivalent to maximizing $\av{W}^\text{out}_\text{fc}= \sum_{i=1}^N Q_i= \Tr H (\rho-\rho')$, Eqs. (\ref{eq:w-fc},\ref{eq:Q_i}). The ergotropy unitaries $\bar U_\alpha$ send each post-selected state $\rho_\alpha$ to its passive companion $\rho'_\alpha=\breve\rho_\alpha=\bar U_\alpha \rho_\alpha \bar U_\alpha^\dagger$, featuring the smallest possible total energy $\Tr H \breve\rho_\alpha$. Thus, their convex combination 
\begin{align}
    \breve \rho=\sum_\alpha q_\alpha \breve\rho_\alpha
\end{align}
features the smallest total energy among all possible non-post-selected states $\rho'$. 
Therefore we have:
\begin{align}
    \eta^L_W(\boldsymbol\Pi)= \eta^L(\boldsymbol\Pi,\bar{\boldsymbol{U}})
\end{align}
which holds regardless of the grain of the measurement.

By setting the cost of information gathering to its minimal Landauer value, Eq. (\ref{eq:Q0L}), the maximum total work output achievable is given by
\begin{align}
W^{\text{max}}_\text{out}= \Tr H (\rho-\breve \rho) ~ -~ h[\{q_\alpha\}]T_0.
\label{eq:max_w-out_general}
\end{align}

For fine-grain measurements, the ergotropy unitaries $\bar{\boldsymbol{U}}$ maximize both the Landauer efficiency and the total work output, that is:
\begin{align}
    \eta^L_W(\boldsymbol{\boldsymbol{F}})= \bar \eta^L(\boldsymbol{F})
\end{align}
where $\boldsymbol{F}$ represents any set of fine-grain measurement projectors. 
In this case, the minimal energy state $\breve \rho$ is nothing but the ground state, and the the maximum total work output  is given by
\begin{align}
W^{\text{max}}_\text{out}= \Tr H (\rho-\ket{e_1}\bra{e_1}) ~ -~ h[\{q_k\}]T_0.
\label{eq:max_w-out_fine}
\end{align}

In general, $\breve\rho$ depends on the operators $\mathbf{\Pi}$, but takes one and the same value $\ket{e_1}\bra{e_1}$ for all fine-grain measurements. As anticipated above, compared to fine-grain measurement, a coarse-grain measurement would exhibit a smaller value of WS output work $\Tr H (\rho-\breve\rho)$.  This, however, is paired with a smaller value of information gathering cost $h[\{q_k\}]T_0$, which gives rise to an intricate cost-benefit trade-off.

\section{Illustrative Examples}
\label{sec:examples}
We now apply our framework to an explicit physical model to analyze the max-work Landauer efficiency $\eta^L_W$ of feedback-controlled quantum heat engines.  Specifically, we consider a WS comprised of two two-level systems (e.g., qubits). The WS Hamiltonian,  Eq. (\ref{eq:H}) is given by 
\begin{align}
H= \frac{\hbar\omega_1}{2}\sigma^z_1 +\frac{\hbar\omega_2}{2}\sigma^z_2
\end{align}
with $\omega_i$ the resonant frequency of qubit $i$ and $\sigma^z_i$ the Pauli-Z operator on qubit $i$.  Furthermore, each qubit is weakly connected to its own thermal bath at temperature $T_i=1/\beta_i$ during the second stroke.

Let the energy eigenstates of qubit $i$ be written as $\ket{0}_i$ and $\ket{1}_i$, which denote the ground and excited states, respectively.  The eigenstates of the two-qubit WS can thus be written as $\ket{a}_1 \otimes \ket{b}_2 = \ket{ab}$, with $a,b \in \{0,1\}$. In the following, all explicit matrices will be written in the computational basis, which coincides with the energy eigenbasis of the system, $\{\ket{00}, \ket{01}, \ket{10}, \ket{11}\}$.  

In the following subsections, we consider using this WS in a feedback-controlled two-stroke quantum heat engine.  We examine the max-work Landauer efficiency $\eta^L_W$ as a function of the bath temperature $T_1$ of qubit 1 in Fig. \ref{fig:eff_vs_T1}. We compare efficiencies for various sets of projective measurements with the blue, red, and purple curves and include for reference the Carnot efficiency with the solid black curve.  We plot the maximal work extracted from the WS $\av{W}_{\text{fc}}^\text{out}=\Tr H(\rho-\breve \rho)$, the minimal cost of information gathering $\av{W}_{\text{m}}=-Q_0^L=h[\{q_k\}]T_0$, and the maximal total work output $W_{\text{out}}$, Eq. (\ref{eq:max_w-out_general}) in Fig. \ref{fig:work_vs_T1}, which give insights into the crossing of curves observed in Fig. \ref{fig:eff_vs_T1}.  For all plots, we use the following fixed parameters: the temperature of the bath connected to qubit 2 $T_2 = 150$ mK, the temperature of the memory bath $T_0 = 80$ mK, and the frequency of both qubits $\omega_i/2\pi = f = 5$ GHz. 

\subsection{Fine-grain measurement in $\boldsymbol{E}$-basis}
We first consider fine-grain measurement of the WS Hamiltonian $H$, defined by the set of projectors in Eq. (\ref{energy_basis}) denoted by $\boldsymbol{E}$. In this case, the four projectors are defined as 
\begin{align}
& E_{00} = \ket{00}\bra{00} ~~~~~~ E_{01} = \ket{01}\bra{01} \nonumber \\
&~~ \nonumber \\
& E_{10} = \ket{10}\bra{10} ~~~~~~ E_{11} = \ket{11}\bra{11} .
\label{eq:E_projectors}
\end{align}
Measurement will collapse the state of the system into one of the states $\rho_\alpha = \ket{\alpha} \bra{\alpha}$, $\alpha \in \{00,01,10,11\}$ with probability $q_\alpha = \bra\alpha \rho \ket \alpha$.  
Explicitly, 
\begin{align}
& q_{00} = \frac{e^{\frac{\beta_1 \hbar \omega_1}{2}}}{Z_1} \frac{e^{\frac{\beta_2 \hbar \omega_2}{2}}}{Z_2} ~~~~~~
q_{01} = \frac{e^{\frac{\beta_1 \hbar \omega_1}{2}}}{Z_1} \frac{e^{\frac{-\beta_2 \hbar \omega_2}{2}}}{Z_2} \nonumber \\
&~~ \nonumber \\
& q_{10} = \frac{e^{\frac{-\beta_1 \hbar \omega_1}{2}}}{Z_1} \frac{e^{\frac{\beta_2 \hbar \omega_2}{2}}}{Z_2} ~~~~~~
q_{11} = \frac{e^{\frac{-\beta_1 \hbar \omega_1}{2}}}{Z_1} \frac{e^{\frac{-\beta_2 \hbar \omega_2}{2}}}{Z_2}
\label{eq:qs}
\end{align}
where $Z_i = \Tr e^{-\beta_i H_i}=2\cosh{(\beta_1 \hbar \omega_i/2)}$ is the partition function of the $i$-th partition of the WS.  
Feedback control consists of applying an associated unitary $U_\alpha$ based on the outcome of the measurement.  To achieve maximum total work output, we select a set of ergotropy unitaries, as defined in Eq. (\ref{eq:ergotropyU}).
Explicitly, we define the feedback unitaries to be:
\begin{equation}
    \begin{aligned}
   & U_{00} =\begin{pmatrix}
    1&0&0&0\\
    0&1&0&0\\
    0&0&1&0\\
    0&0&0&1\\
    \end{pmatrix} ~~~~~
    U_{01} =\begin{pmatrix}
    0&1&0&0\\
    1&0&0&0\\
    0&0&1&0\\
    0&0&0&1\\
    \end{pmatrix} \\
    & \\
    & U_{10} =\begin{pmatrix}
    0&0&1&0\\
    0&1&0&0\\
    1&0&0&0\\
    0&0&0&1\\
    \end{pmatrix}  ~~~~~
    U_{11} =\begin{pmatrix}
    0&0&0&1\\
    0&1&0&0\\
    0&0&1&0\\
    1&0&0&0\\
    \end{pmatrix} 
    \label{eq:feedback_unitaries}
    \end{aligned}
\end{equation}
Note that each $U_\alpha$ is simply a permutation between the energy eigenstate $\ket{\alpha}$ and the ground state (i.e., the lowest energy eigenstate).  

The final state after measurement and application of the associated feedback unitary is $\rho'_\alpha = U_\alpha \rho_\alpha U_\alpha^\dagger = \ket{00} ~ \forall ~ \alpha$.  The non-post-selected final state is $\rho' = \sum_\alpha q_\alpha \ket{00}  = \ket{00}$. The average change in energy of each qubit $i$ can thus be computed with Eq. (\ref{eq:delta_E}), plugging in the ground state for $\rho'$.  Since $\rho'$ is the ground state, we will have $\av{\Delta E_i} \le 0$ for both qubits, and thus $Q_i \ge 0$ for both qubits, according to Eq. (\ref{eq:Q_i}).  Therefore, $Q_{in} = \sum_{i=1}^2 Q_i = -\sum_{i=1}^2 \av{\Delta E_i}$.  

The max-work Landauer efficiency $\eta^L_W$ for fine-grain measurement in the $\boldsymbol{E}$-basis is shown by the dotted blue curve in Fig. \ref{fig:eff_vs_T1}.  Likewise, the dotted blue curves in Fig. \ref{fig:work_vs_T1} plot the average extracted work (a), the cost of information gathering (b), and the total work output (c) for this measurement.  These will be compared with various other projective measurements in the following subsections.

\subsection{Coarse-grain measurement in $\boldsymbol{E}$-basis}
We next consider a coarse-grain measurement in the $\boldsymbol{E}$-basis.  There are an infinite number of ways to perform a coarse-grained measurement, but we select an intuitive one here for explicit illustration.  

We define the coarse-grain measurement projectors as 
\begin{align}
\Pi_1 = E_{00}; ~~ \Pi_2 = E_{01} + E_{10}; ~~ \Pi_3 = E_{11}.
\end{align}
In words, this measures how many excitations are present in the system (either 0, 1, or 2). Measurement will produce the states 
\begin{align}
   & \rho^C_1 =\begin{pmatrix}
    1&0&0&0\\
    0&0&0&0\\
    0&0&0&0\\
    0&0&0&0\\
    \end{pmatrix}
    ~~~~
    \rho^C_2 = \frac{1}{q_{01} + q_{10}}\begin{pmatrix}
    0&0&0&0\\
    0&q_{01}&0&0\\
    0&0&q_{10}&0\\
    0&0&0&0\\
    \end{pmatrix}
    \nonumber \\
    & \rho^C_3 =\begin{pmatrix}
    0&0&0&0\\
    0&0&0&0\\
    0&0&0&0\\
    0&0&0&1\\
    \end{pmatrix}
\end{align}
with associated probabilities $p_1 = q_{00}$, $p_2 = q_{01} + q_{10}$, and $p_3 = q_{11}$, respectively, where the $q_\alpha$ are defined in Eq. (\ref{eq:qs}).  

Again, we select a set of ergotropy unitaries to maximize total work output. Following Ref. \cite{Allahverdyan04EPL67}, these can be constructed as unitaries which sort the eigenvalues of the corresponding post-measurement density matrix of the WS in non-increasing order. We define unitaries $U_1 = U_{00}$ and $U_3 = U_{11}$ are as previously defined in the fine-grain case.  The feedback unitary $U^C_2$ associated to $\Pi_2$ is written explicitly as
\begin{align}
U^{C}_2 = \begin{cases} & \begin{pmatrix}
    0&0&1&0\\
    0&1&0&0\\
    1&0&0&0\\
    0&0&0&1\\
    \end{pmatrix} \text{if} ~~ q_{10} > q_{01}
     \\
     \\
  & \begin{pmatrix}
    0&1&0&0\\
    0&0&1&0\\
    1&0&0&0\\
    0&0&0&1\\
    \end{pmatrix} \text{if} ~~ q_{01} > q_{10}.
    \end{cases}
\end{align}
Since the $q_\alpha$ are determined by the initial parameters of the WS, we know at the outset of the problem how to define $U_2^C$. Note that in the case of coarse-grain measurement, the final state after measurement and feedback $\rho'$ will not always be the pure ground state, but rather a mixture of states, which reduces the work extracted from the WS, and thus, the total work output.  

\begin{figure}\label{fig:eff_vs_T1}
\centering
\begin{overpic}[width=0.94\columnwidth]{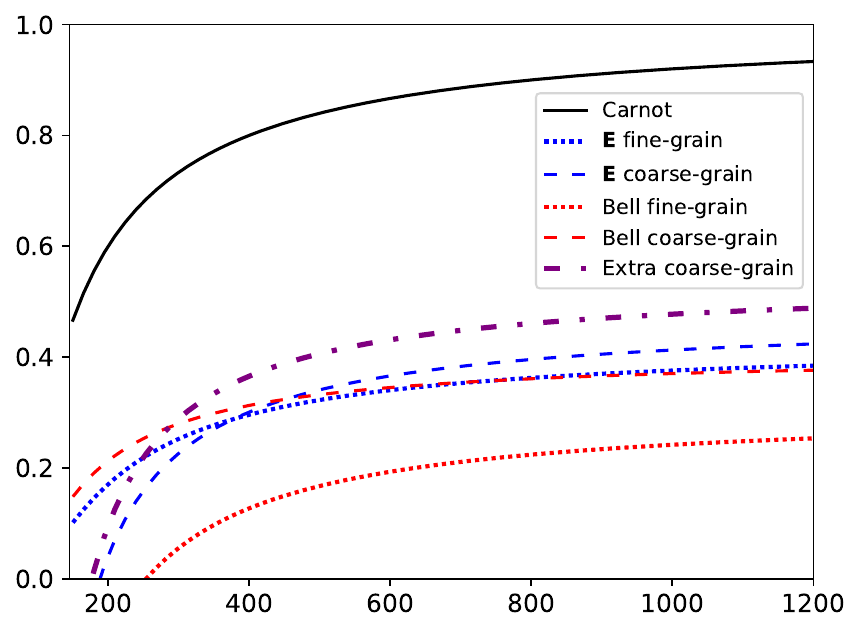}
\put(-3,36){\rotatebox{90}{$\eta^L_W$}}
\put(50,-2){$T_1$ [mK]}
\end{overpic}
\caption{Comparison of max-work Landauer efficiency $\eta^L_W$ for fine- and coarse-grain measurement in the $\boldsymbol{E}$ and Bell bases as a function of bath temperature $T_1$ of qubit 1.}
\end{figure}

The max-work Landauer efficiency $\eta^L_W$ when using this coarse-grain measurement is plotted by the dashed blue curve in Fig. \ref{fig:eff_vs_T1}. We compare this with fine-grain measurement in the same basis (dotted blue curve), which shows there is a cross-over point in efficiency.  This highlights the intriguing observation that more information is not always better.  Indeed, there is a real cost to gathering information from a system, and given a particular set of initial parameters, it may be more efficient to gather less information.

Figure \ref{fig:work_vs_T1} provides some insight into the cross-over in max-work Landauer efficiencies between the fine- and coarse-grain measurements.  Figure \ref{fig:work_vs_T1}c shows that the total work output is only slightly greater in the fine-grain case (dotted blue curve) versus the coarse-grain case (dashed blue curve) for measurement in the $\boldsymbol{E}$ basis.  However, there is a larger gap between the two cases for the cost of gathering information, Figure \ref{fig:work_vs_T1}b.  While the efficiency in the fine-grain case is augmented with greater work output, it is hampered by the larger cost of gathering information.  Thus, when the cost of information gathering is not well compensated by its corresponding improvement in work output, coarse-grained measurements may have an advantage over fine-grained measurement in terms of efficiency.

\subsection{Fine-grain measurement in Bell basis}
\label{sec:Bell_fine-grain}
We next turn to performing measurements in the Bell basis, given by 
\begin{align}
\ket{\Phi^{\pm}} = \frac{1}{\sqrt2}(\ket{00} \pm \ket{11}) ~~~~
\ket{\Psi^\pm} = \frac{1}{\sqrt2}(\ket{01} \pm \ket{10}).
\label{eq:bell_basis}
\end{align}
We consider a fine-grain measurement in this basis defined by the four projectors 
\begin{align}
&\Pi^B_1 = \ket{\Phi^+}\bra{\Phi^+} ~~~~~ \Pi^B_2 = \ket{\Psi^+}\bra{\Psi^+} \nonumber \\
& \Pi^B_3 = \ket{\Phi^-}\bra{\Phi^-} ~~~~~ \Pi^B_4 = \ket{\Psi^-}\bra{\Psi^-}.
\label{eq:bell_projectors}
\end{align}
Measurement will collapse the state of the system into one of the Bell states $\Pi^B_\alpha$ with probability $p^B_\alpha = \Tr \rho \Pi^B_\alpha$. Explicitly, $p^B_1 = p^B_3 = \frac{1}{2}(q_{00} + q_{11})$ and $p^B_2 = p^B_4 = \frac{1}{2}(q_{01} + q_{10})$, where the $q_\alpha$ were defined in Eq. (\ref{eq:qs}). 

One way to determine the ergotropy unitaries for maximal total work output is as follows. First, define a unitary transformation $U_t$ that rotates each Bell state to a unique eigenstate of the Hamiltonian $H$.  Applying $U_t$ to the measured Bell state will transform the state to its corresponding eigenstate of $H$.  Next, the appropriate unitary $U_\alpha$ from Eq. (\ref{eq:feedback_unitaries}) can be applied to transform the state to the ground state of the system.  Thus, the feedback unitary for each outcome of the fine-grain Bell-basis measurement can be derived by composing $U_t$ with the associated $U_\alpha$.  

Explicitly, we define $U_t$ as follows: 
\begin{equation}
   U_t = \frac{1}{\sqrt{2}}\begin{pmatrix}
    1&0&0&1\\
    0&1&1&0\\
    1&0&0&-1\\
    0&1&-1&0\\
    \end{pmatrix}
\end{equation}
which leads to the following feedback unitaries for maximal work output:
\begin{align}
&U^B_1 = U_t U_{00} ~~~~~ U^B_2 = U_t U_{01} \nonumber \\
& U^B_3 = U_t U_{10} ~~~~~ U^B_4 = U_t U_{11}.
\label{eq:fine_Bell_Us}
\end{align}

All feedback unitaries $U^B_i$ map the post-measurement state to the pure ground state. Thus, as in the case for fine-grain measurement in the $\boldsymbol{E}$-basis, the final state after measurement and feedback is $\rho'_\alpha = \ket{00}\bra{00} ~ \forall ~ \alpha$.  
\begin{figure*}\label{fig:work_vs_T1}
\centering
\begin{overpic}[width=2\columnwidth]{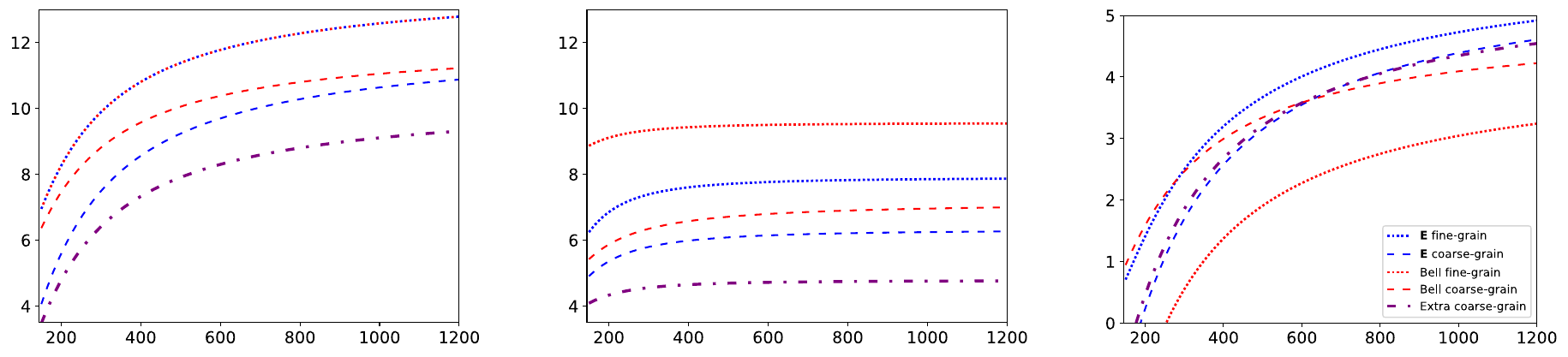}
\put(-2,24){(a)}
\put(32,24){(b)}
\put(67,24){(c)}
\put(-2,7){\rotatebox{90}{$\av{W}^{\text{out}}_{\text{fc}} ~~ [\mu eV]$}}
\put(33,8){\rotatebox{90}{$\av{W}_{\text{m}} ~~ [\mu eV]$ }}
\put(68,9){\rotatebox{90}{$W_{\text{out} } ~~ [\mu eV]$ }}
\put(14,-1){$T_1$ [mK]}
\put(49,-1){$T_1$ [mK]}
\put(83,-1){$T_1$ [mK]}
\end{overpic}
\caption{Comparison of (a) feedback control work extraction, $\av{W}^{\text{out}}_{\text{fc}}$, (b) Landauer cost of information gathering $\av{W}_{\text{m}}$, and (c) total work output $W_{\text{out}}$, for fine- and coarse-grain measurements in the $\boldsymbol{E}$ and Bell bases as a function of bath temperature $T_1$ of qubit 1.}
\end{figure*}

The max-work Landauer efficiency $\eta^L_W$ for fine-grain measurement in the Bell basis is shown by the dotted red curve in Fig. \ref{fig:eff_vs_T1}. As expected for fine-grain measurements, $\eta^L_W$ is strictly smaller in the Bell basis (or in any other basis) than in the $\boldsymbol{E}$-basis, since we showed that the $\boldsymbol{E}$-basis has optimal efficiency in the fine-grain measurement setting. Fig. \ref{fig:work_vs_T1}c shows that total work output for fine-grain Bell measurement is the lowest of all measurements considered, while Fig. \ref{fig:work_vs_T1}b shows that the cost of information gathering is the highest.  These both negatively impact the efficiency, explaining why coarse-grain measurement in the Bell basis leads to the lowest efficiency of all measurements considered, as shown in Figure \ref{fig:eff_vs_T1}.

\subsection{Coarse-grain measurement in the Bell basis}
We next turn to coarse-grained measurement in the Bell basis.  We select a coarse-graining that is analogous to the coarse-grain measurement in the $\boldsymbol{E}$ basis (i.e., two rank-1 projectors and one rank-2 projector), defined by the following set of projectors
\begin{align}
\Pi^{CB}_1 = \Pi^{B}_2 ; ~~~~ 
\Pi^{CB}_2 = \Pi^{B}_4 ; ~~~~ 
\Pi^{CB}_3 = \Pi^{B}_1 + \Pi^{B}_3.
\label{eq:coarse_bell_projs}
\end{align}

Measurement will collapse the state of the system into the states 
\begin{align}
   &\rho^{CB}_1 = \Pi^B_2, ~~~~ 
   \rho^{CB}_2 = \Pi^B_4, ~~~~ \nonumber \\
   &\rho^{CB}_3 = \frac{1}{q_{00} + q_{11}} \begin{pmatrix}
    q_{00}&0&0&0\\
    0&0&0&0\\
    0&0&0&0\\
    0&0&0&q_{11}\\
    \end{pmatrix}
\end{align}
where the $q_\alpha$ are defined in Eq. (\ref{eq:qs}), with associated probabilities $p^{CB}_1 = p^B_2$, $p^{CB}_2 = p^B_4$, $p^{CB}_3 = p^B_1 + p^B_3$, with $p^B_i$ defined in subsection \ref{sec:Bell_fine-grain}. 
We again select ergotropy unitaries to maximize total work output, written explicitly as
\begin{align}
   U^{CB}_1 = U^B_2; ~~~~
   U^{CB}_2 = U^B_4; ~~~~
   U^{CB}_3 = \begin{pmatrix}
    1&0&0&0\\
    0&0&0&1\\
    0&0&1&0\\
    0&1&0&0\\
    \end{pmatrix}
\end{align}
where $U^B_{i}$ were defined in Eq. \ref{eq:fine_Bell_Us}.

The max-work Landauer efficiency $\eta^L_W$ for coarse-grain measurement in the Bell basis is shown by the dashed red curve in Fig. \ref{fig:eff_vs_T1}.  We observe that a significantly higher efficiency can be achieved with a coarse-grained measurement in the Bell basis as compared to a fine-grain measurement in the same basis. This is corroborated in Fig. \ref{fig:work_vs_T1}, which shows how the total work output is larger and the information gathering cost is smaller in the coarse-grain measurement compared to the fine-grain measurement in the Bell basis, both of which contribute to the higher efficiency for the coarse-grain measurement.

\subsection{Extra-coarse-grained measurement}
We finally turn to an even further coarse-grained measurement.  While the previous two examples with coarse-graining featured projector sets with two rank-1 projectors and one rank-2 projector, we now examine an extra-coarse-grained measurement featuring two rank-2 projectors. The projectors can be written in either the $\boldsymbol{E}$ or Bell bases as follows:
\begin{align}
&\Pi^{E}_1 = \Pi^{B}_1 + \Pi^{B}_3 = E_{00} + E_{11} \\
&\Pi^{E}_2 = \Pi^{B}_2 + \Pi^{B}_4 = E_{01} + E_{10}.
\label{eq:coarse_bell_projs}
\end{align}
In words, this observable measures whether the qubits are in the same state (i.e., both in the ground state or both in the excited state) or in opposite states (i.e., one qubit in ground state, one qubit in excited state). Measurement will collapse the state of the system into the states 
\begin{align}
   &\rho^{E}_1 = \frac{1}{q_{00} + q_{11}} \begin{pmatrix}
    q_{00}&0&0&0\\
    0&0&0&0\\
    0&0&0&0\\
    0&0&0&q_{11}\\
    \end{pmatrix}
    \nonumber \\
    \nonumber \\
    &\rho^{E}_2 = \frac{1}{q_{01} + q_{10}} \begin{pmatrix}
    0&0&0&0\\
    0&q_{01}&0&0\\
    0&0&q_{10}&0\\
    0&0&0&0\\
    \end{pmatrix}
\end{align}
with probabilities $p^{E}_1 = q_{00} + q_{11}$ and $p^{E}_2 = q_{01} + q_{10}$, respectively, where the $q_\alpha$ are defined in Eq. (\ref{eq:qs}).  We select ergotropy unitaries to maximize total work output.  Explicitly, the feedback unitaries are defined as
\begin{align}
   U^{E}_1 = \begin{pmatrix}
    1&0&0&0\\
    0&0&0&1\\
    0&0&1&0\\
    0&1&0&0\\
    \end{pmatrix}
    ~~
   U^{E}_2 = \begin{cases} & \begin{pmatrix}
    0&0&1&0\\
    0&1&0&0\\
    1&0&0&0\\
    0&0&0&1\\
    \end{pmatrix} \text{if} ~~ q_{10} > q_{01}
     \\
     \\
  & \begin{pmatrix}
    0&1&0&0\\
    0&0&1&0\\
    1&0&0&0\\
    0&0&0&1\\
    \end{pmatrix} \text{if} ~~ q_{01} > q_{10}.
    \end{cases}
\end{align}
Since the $q_\alpha$ are determined by the initial parameters of the WS, $U^{E}_2$ can be defined at the outset of the problem. 

We plot $\eta^L_W$ for the extra coarse-grain measurement in Figure \ref{fig:eff_vs_T1} with the dash-dot purple curve. It features interesting relationships with the efficiencies of the other measurements we have considered.  Comparing the efficiency of the extra coarse-grain measurement with those of the $\boldsymbol{E}$ basis measurements, we observe that the efficiency of the extra coarse-grain measurement is strictly greater than that of the coarse-grain measurement, but experiences a cross-over with that of the fine-grain measurement.  Comparing the efficiency of the extra coarse-grain measurement with those of the Bell basis measurements, we observe the opposite.  Namely, the efficiency of the extra coarse-grain measurement is strictly greater than that of the fine-grain measurement, but experiences a cross-over with that of the coarse-grain measurement.


The most remarkable observation emerging from Fig. \ref{fig:work_vs_T1} is that there are crossing points in the total work output curves between the coarse and extra-coarse measurements in both bases. This reveals the intriguing fact that gathering less information will in some instances not only achieve a higher efficiency, but also achieve a higher total work output. Intuitively, one might expect that gathering less information would result in a lower total work output, since information generally aids in extracting more work.  However, the total work output, Eq. (\ref{eq:W-out}), contains two competing terms.
The first term, $\av{W}^\text{out}_\text{fc}$, decreases with increasing coarseness (i.e., less information), thereby \textit{decreasing} total work output.  However, the second term, $\av{W}_\text{m}$, also decreases with increasing coarseness, thereby \textit{increasing} total work output.  Our plots demonstrate that depending on the system parameters, increasing coarseness may cause the benefit of decreasing the latter term to outweigh the disadvantage of decreasing the former term, thus resulting in an overall advantage (i.e., higher total work output). Interestingly, Fig. \ref{fig:eff_vs_T1} and Fig. \ref{fig:work_vs_T1}c demonstrate that there are ranges of system parameters where extra coarse-grain measurement achieves both better total work output and efficiency compared to a moderate coarse grain measurement.  In other words, in some scenarios, less information can be advantageous for both total work output and efficiency.

\section{Discussion}
Before concluding, there are a few remarks that warrant discussion.  First, while we have focused on projective measurements, the theory can be extended to include generic POVMs. In that case the proof of Eq. (\ref{h>DH}) would be more involved, and can be achieved by following the lines of Sagawa and Ueda \cite{Sagawa08PRL100}.

Second, most of our results have relied on the assumption that the thermodynamic cost incurred in the measurement process saturates the Landauer bound, Eq. (\ref{eq:av-m>h}).  However, in an experimental realization, one typically expects that the actual cost will be much higher. This is because the measurement involves the displacement of a macroscopic pointer, whose associated work would typically scale with its mass, i.e., in an extensive way. See for example \cite{Nieuwenhuizen25FQST4} where this was observed in the Curie-Weiss model of quantum measurement, where the state of a spin is imprinted in the total magnetization of a ferromagnet \cite{Allahverdyan03EPL61,Allahverdyan13PR525}. This is also corroborated by Ref. \cite{Buffoni23QUANTUM7}, where the encoding of a single bit into the magnetization of a macroscopic Ising network of $N$ superconducting qubits, was experimentally observed to have a cost of the order $N k T$. See also the numerical study of Ref. \cite{Buffoni22JSP186}. This evidences that, in practice, the positive energy output achieved by the feedback control $\av{W}^\text{out}_\text{fc}$, could be greatly overcome by the positive energetic cost of the measurement $\av{W}_\text{m}$, see Eq (\ref{eq:W-out}).

In order to keep the thermodynamic balance positive, it is crucial to approach the Landauer bound.  However, as highlighted in Refs. \cite{Guryanova20QUANTUM4,Taranto23PRXQ4}, this comes at the cost of other resources, namely, either the time duration or the complexity of the measurement protocol.  In order for the machine to be useful, the total time duration of the cycle should be kept finite in order to output some finite power.  Thus, the cost must likely be paid in terms of complexity of the measurement protocol \cite{Guryanova20QUANTUM4}.


Third, we note that besides the time scale of measurement $\tau_m$, there are two more relevant time scales in our theory: the time scale $\tau_u$ of the unitary driving in the first stroke, and the thermal relaxation time scale $\tau_r$ in the second stroke. The relaxation time $\tau_r$ can greatly vary from system to system, but the challenge is typically  that of making it long, rather than making it short. The unitary operation time scale $\tau_u$ can similarly greatly vary from case to case. Generally, one can trade it with increased complexity of the unitary or with decreased energy output $\av{W}_\text{fc}^\text{out}$ \cite{Taranto23PRXQ4,Oftelie24PRXQ5}. These two time scales are present as well in standard feedbackless two-stroke quantum heat engines \cite{Quan07PRE76,Campisi15NJP17}.

Finally, we note that we have focused only on the heat engine operation regime in this work. The cooling operation regime, which is characterized by extraction of heat from the coldest bath, is also of great interest for various applications. We leave analysis of its maximal efficiency and heat extraction, along with their dependence on the choice of projectors $\mathbf{\Pi}$, for future work. Due to the high non-linearity of the problem, we foresee a plethora of interesting phenomena and trade-offs occurring in this regime as well.

\section{Conclusion}
By grounding our analysis of feedback-controlled quantum thermal machines in the explicit physical model of a two-stroke heat engine, we were able to uncover intriguing observations about the intricacy of the effect information has on the efficiency of such machines.  Information has been demonstrated to augment performance of quantum engines, sometimes even endowing functionality to a system that would otherwise provide no useful output (as in the case of the Szilard engine).  One might therefore conclude that more information always leads to better performance in feedback controlled machines.  However, upon analysis of explicit scenarios, we found that maximum information extraction can lead to lower efficiency, which may at first seem counter-intuitive.  

By placing the WS and the demon's memory on equal footing, as we do in our approach, this observation becomes more apparent in hindsight.  The extracted information must be written into a memory, which is stored at some temperature $T_0$.  There is an associated, unavoidable, work cost with storing this information, which will scale with $T_0$.  As $T_0$ is independent of the bath temperatures associated with the WS, there are situations in which the cost of storing more information can outweigh the benefits of the addition work extraction this information provides.  In these instances, efficiency of the engine can be improved by extracting less information (i.e., coarse-graining the measurement).  

It is interesting to consider the implications of such an observation.  For example, consider a very rough description for a living organism, which at a fundamental level performs measurements and decides its next action based on the information collected from the measurement.  A simplified model for an organism could thus be viewed as a feedback-controlled engine.  For living organisms, a key metric for survival is efficiency, as energy is a precious resource.  The type of measurements an organism evolves to performs should therefore maximize the efficiency.  Our results indicate that these measurements are thus not necessarily those which extract maximal information.  Coarse-graining information intake may indeed be instrumental for living organisms' survival.

Bringing the discussion back to more well-defined systems, our results have more direct and immediate applicability to design considerations for miniature devices featuring quantum thermal machines.  When constructing such devices, there is not only room to optimize engine performance based on the parameters of the WS itself, but also based on the amount of information gained from measurements, which can be tuned based on the type of measurement performed (i.e., the measurement basis used, and the level of coarse-graining).  

\begin{acknowledgments}
LBO gratefully acknowledges funding from the European Union’s Horizon 2020 research and innovation program under the Marie Skłodowska-Curie grant agreement No 101063316.
\end{acknowledgments}


\bibliography{references}

\newpage
\appendix

\section{Proof of Eq. (\ref{h>DH})}
\label{sec:proof}
Since von Neumann information is convex and invariant under unitary transformations, using Eqs. (\ref{eq:rho'alpha},\ref{eq:rho'}), we have
\begin{align}
\mathcal H[\rho'] \geq  \sum_{\alpha=1}^K q_\alpha \mathcal H[\rho'_\alpha] =   \sum_{\alpha=1}^K q_\alpha \mathcal H[\rho_\alpha].
\end{align}
Let us consider the density operator 
\begin{align}
\bar{\rho}= \sum_{\alpha=1}^K  q_\alpha \rho_\alpha =  \sum_{\alpha=1}^K \Pi_\alpha \rho \Pi_\alpha.
\label{eq:barrho}
\end{align}
Let $\ket{\psi_k}$ be its eigenvectors, and $r_k$ the corresponding eigenvalues. Since $\bar{\rho} $ is block-diagonal with each block represented by $q_\alpha \rho_\alpha$, its eigenbasis $\{\ket{\psi_k}, k = 1\dots d\}$  is the union of the eigenbases  $\{\ket{\psi_k}, k \in I_{\alpha}\}$ of the operators $q_\alpha \rho_\alpha$, spanning the according subspaces, labelled by the index $\alpha$. Here $I_\alpha$ is the set of labels $k$ referring to the eigenstates $\ket{\psi_k}$ spanning the $\alpha$ subspace. Note that the number of elements of $I_\alpha$ is $g_\alpha$, and
$\cup_\alpha I_\alpha = \{1 \dots d \} $. 

The operator $\rho_\alpha$ is diagonal in the basis $\{\ket{\psi_k}, k \in I_{\alpha}\}$ with eigenvalues $r_k/q_\alpha$. Thus,
\begin{align}
 \mathcal H[\rho_\alpha] = -\sum_{k \in I_\alpha} (r_k/q_\alpha) \ln  (r_k/q_\alpha) 
\end{align}
therefore:
\begin{align}
 \sum_{\alpha=1}^K q_\alpha \mathcal H[\rho_\alpha] 
= -  \sum_{\alpha=1}^K \sum_{k \in I_\alpha} r_k \ln r_k +   \sum_{\alpha=1}^K \sum_{k \in I_\alpha} r_k \ln  q_\alpha
\end{align}
Furthermore, $\sum_{k \in I_\alpha} r_k = \Tr \Pi_\alpha \rho \Pi_\alpha= q_\alpha$, and  $ \sum_{\alpha=1}^K \sum_{k \in I_\alpha}=\sum_{k=1}^d$, thus:
\begin{align}
 \sum_{\alpha=1}^K q_\alpha \mathcal H[\rho_\alpha] 
&= -\sum_{k=1}^d r_k \ln r_k +   \sum_{\alpha=1}^K q_\alpha  \ln  q_\alpha \nonumber \\
&
=\mathcal H[\bar{\rho}] - h[\{q_\alpha\}]
\end{align}
Finally, note that the linear transformation that maps $\rho$ onto $\bar \rho$, Eq. (\ref{eq:barrho}), is unital. Therefore $\mathcal H[\bar{\rho}] \geq \mathcal H[\rho]$ \cite{Sagawa20Book}.
Combining everything together we get
\begin{align}
\mathcal H[\rho'] &\geq \mathcal H[\bar{\rho}] - h[\{q_\alpha\}]\geq \mathcal H[\rho] - h[\{q_\alpha\}] 
\end{align}
which proves Eq. (\ref{h>DH}).

\newpage
\end{document}